\documentclass[twoside]{article}
\usepackage{fleqn,espcrc2}

\usepackage{graphicx,epsfig}
\usepackage[figuresright]{rotating}

\newcommand{\AmS}{{\protect\the\textfont2
  A\kern-.1667em\lower.5ex\hbox{M}\kern-.125emS}}

\newcommand{\nn}{\nonumber}
\newcommand{\raw}{\rightarrow}

\newcommand{\bea}{\begin{eqnarray}}
\newcommand{\eea}{\end{eqnarray}}

\def\beq{\begin{equation}}
\def\eeq{\end{equation}}

\newcommand{\tetaot}{\mbox{$\theta_{13}$}}
\newcommand{\tetatt}{\mbox{$\theta_{23}$}}
\newcommand{\deltt}{\mbox{$\Delta_{23}$}}
\newcommand{\delot}{\mbox{$\Delta_{13}$}}

\newcommand{\numu}{\mbox{$\nu_{\mu}$}}
\newcommand{\numubar}{\mbox{$\overline{\nu}_{\mu}$}}
\newcommand{\nue}{\mbox{$\nu_{e}$}}

\newcommand{\muplus}{\mbox{$\mu^{+}$}}

\title{ Measurement of CP violation at a Neutrino Factory}

\author{J.J. G\'omez-Cadenas \address{IFIC, Edificio de Institutos 
de Paterna, Apto Correos 2085, 46071 Valencia, Spain and 
CERN, CH-1211 Geneve 23, Switzerland}} 
\begin{document}

\begin{abstract}
The prospects of measuring CP violation in the leptonic sector
using the intense neutrino beams arising from muon 
decay in the straight sections of a muon accumulator ring
(the so-called {\em neutrino factory}) are discussed.  
\vspace{1pc}
\end{abstract}

\maketitle

\section{INTRODUCTION}
\label{intro}
In this paper I discuss the prospects to observe a CP-odd phase in the leptonic
sector, using the intense, pure two-flavour neutrino beams produced in a future
neutrino facility, the so-called {\em neutrino factory}.

The organization is as follows. Section \ref{ref:nuosc} summarizes the state-of-the-art
concerning neutrino oscillations. Section \ref{ref:nufact} describes the neutrino
factory. The prospects to measure a CP-odd
phase are discussed in section \ref{ref:cp}.
\section{NEUTRINO OSCILLATIONS}
\label{ref:nuosc}

Perhaps the most exciting physics result 
of the last two decades in the field of particle physics\footnote{Excluding, of course, the possible discovery of a Higgs particle by the LEP experiments.} is the
growing evidence that neutrinos have mass and oscillate. There are three independent
sets of experimental data that support this hypothesis. They are:
\begin{enumerate}
\item The measurements of the rates (both absolute and as a function of
the zenith angle) of atmospheric neutrinos by the experiment SuperKamiokande
and others\cite{Superka,otheratm}. 
The observed $\nu_\mu$ rate is about 50 \% smaller,
while the observed $\nu_e$ rate is consistent with the predicted rate. In
addition, the rate reduction of the observed $\nu_\mu$ varies with the
incoming neutrino zenith angle 
as expected if oscillations are at play. Recent analysis of the atmospheric
data\cite{newSK,concha} favor strongly
oscillations of $\nu_\mu$'s into $\nu_\tau$'s, while almost completely excluding
oscillations into sterile neutrinos. The mass gap\footnote{ 
$\Delta m^2_{ij} \equiv m^2_j - m^2_i$.} between the two oscillating neutrinos,
$\Delta m^2_{23} = \Delta m^2_{atm}$ is in the range $10^{-3}$--$10^{-2} eV^2$, while
the mixing angle $\theta_{23} = \theta_{atm}$ is 
close to maximal.
\item The measurement of the rates of solar neutrinos, by several 
experiments\cite{solar}. 
The solar neutrino deficit is interpreted either as MSW (matter
enhanced oscillations)\cite{MSW} or as vacuum oscillations (VO)\cite{osc} 
that deplete the original $\nu_e$'s  presumably in favor of $\nu_\mu$'s 
(oscillation into sterile neutrinos are also disfavored\cite{newSK,concha}). 
The corresponding squared mass differences are:
{\em (i)} $ \Delta m^2_{12} = \Delta m^2_{sun} \sim 10^{-5}-10^{-4}~eV^2$ 
for the large mixing angle MSW solution (LMA-MSW); 
{\em (ii)} $ \Delta m^2_{12} \sim 10^{-6} eV^2$ for the 
small mixing angle MSW solution (SMA-MSW) and
{\em (iii)}  $\Delta m^2_{12} \sim 10^{-10} eV^2$ for VO. The mixing angle
is close to maximal for both the LMA-MSW solution and the VO solution 
and small ($\sin^2 2 \theta_{12} = \sin^2 2 \theta_{sun} \sim 10^{-3}$) for
the SMA-MSW solution.
\item The evidence of neutrino oscillations claimed by 
the LSND collaboration\cite{lsnd}. This experiment has operated
in an almost-pure $\nu_\mu$ beam, and observes an excess of $\nu_e$'s over
their calculated background. They interpret their results in terms
of oscillations of $\nu_\mu$'s into $\nu_e$'s, with a squared mass difference
$\Delta m^2_{lsnd} \sim 1 eV^2$.
\end{enumerate}

One obvious fact that follows from the existence of three different mass squared
differences, $\Delta m^2_{sun} << \Delta m^2_{atm} << \Delta m^2_{lsnd}$ is 
that more than three neutrinos are needed in order to explain all data simultaneously.
This would require sterile neutrinos, which are disfavored by current experimental
data\cite{newSK}.
Alternatively, to explain oscillations with
three standard neutrinos one must discard some of the data. This will be,
apologetically, my approach in this paper.\footnote{It is Currently fashionable 
to disbelieve the LSND results. However, they have not
been proved wrong, so far, by an alternative experiment. One such experiment, MiniBoone is
approved in FNAL.} I will consider only the two strongest evidences for neutrino oscillations,
namely, the solar and atmospheric anomalies and I will, for simplicity 
assume Dirac neutrinos. Under this assumptions
the NMS matrix\footnote{The NMS matrix is the equivalent to the CKM matrix in the lepton sector.}
connecting the flavor and mass eigenstates, 
$(\nu_e,\nu_\mu,\nu_\tau)^T= U_{NMS} \cdot (\nu_1,\nu_2,\nu_3)^T$, 
contains four physical parameters, i.e., three mixing angles and a CP-odd phase,
 and can be conveniently parameterized as:
\begin{eqnarray}
U && \equiv  U_{23} U_{13} U_{12} 
\equiv
\left(\matrix{         1 &        0 &       0              \cr 
                       0 &   c_{23} &  s_{23}              \cr
                       0 & - s_{23} &  c_{23}              \cr } \right) \\ \nonumber
& & 
\left(\matrix{    c_{13} &        0 &  s_{13} e^{i \delta} \cr
                       0 &        1 &       0              \cr
 -  s_{13} e^{-i \delta} &        0 &  c_{13}              \cr } \right) 
\left(\matrix{    c_{12} &   s_{12} &       0              \cr
                - s_{12} &   c_{12} &       0              \cr
                       0 &        0 &       1              \cr } \right) 
\label{CKM}
\end{eqnarray}
with $s_{12} \equiv \sin \theta_{12}$, and similarly for the other 
sines and cosines.

Neutrino oscillations are due to the fact that neutrinos produced in 
a weak eigenstate can change flavor as they propagate a distance ~$L$
from the production point.
 
In Vacuum, defining 
the product of NMS matrix elements
$W_{\alpha\beta}^{jk}\equiv  \,[V_{\alpha j}V_{\beta j}^* V_{\alpha k}^*V_{\beta k}]$,
one can write the probability of a neutrino (antineutrino) of flavor $\alpha$ to
oscillate into a neutrino (antineutrino) of flavor $\beta$ as:

\begin{eqnarray}
P(& &\nu(\bar \nu)_\alpha  \rightarrow  \nu(\bar \nu)_\beta)  =  \nonumber \\ 
& & -4\; \sum_{k>j}\,\rm{Re}[W_{\alpha\beta}^{jk}]\, 
\sin^2\left({\Delta m^2_{jk}\,L\over 4 E_\nu}\right)
\,\nonumber\\ 
&\pm &\, 2  \,
\sum_{k>j}\, \rm{Im}[W_{\alpha\beta}^{jk}]\, \sin\left({\Delta m^2_{jk}\,L \over 2 E_\nu}\right)
\label{eq:prob}
\end{eqnarray}

Equation \ref{eq:prob} contains a CP-even 
($-4\; \sum_{k>j}\,\rm{Re}[W_{\alpha\beta}^{jk}]\, 
\sin^2\left({\Delta m^2_{jk}\,L\over 4 E_\nu}\right)$) 
and a CP-odd 
($\sum_{k>j}\, \rm{Im}[W_{\alpha\beta}^{jk}]\, \sin\left({\Delta m^2_{jk}\,L \over 2 E_\nu}\right)$) 
term, which is only different
from zero if there is at least an imaginary phase in the NMS matrix. This is, of course the
case for three families, but not for two families. In this case the oscillation reduces to 
the familiar formula:   

\begin{eqnarray}
P_{\nu_\alpha\nu_\beta} =  \sin^2 2\; \theta \; \sin^2\left({\Delta m^2\,L\over 4 E_\nu}\right) 
\label{eq:twof}
\end{eqnarray}

On the other hand, the fact that $\Delta m^2_{atm} >> \Delta m^2_{sun}$ permits to describe
accurately neutrino oscillation probabilities at terrestrial distances with
only three parameters, $\tetatt$, $\Delta m_{23}^2 = \Delta m_{atm}^2$ and $\tetaot$:
Equation \ref{eq:prob} then simplifies to:
\begin{eqnarray}
P_{\nu_e\nu_\mu} & = & \;  \sin^2 2 \;\theta_{13}  \sin^2 \theta_{23} \;
 \sin^2 \frac{\Delta m^2_{23} L}{4 E_\nu} \nonumber\\
P_{\nu_e\nu_\tau} & = & \; \sin^2 2 \; \theta_{13}
 \cos^2 \theta_{23}   \;  \sin^2 \frac{\Delta m^2_{23} L}{4 E_\nu} \nonumber\\
P_{\nu_\mu\nu_\tau} & = & \;  \sin^2 2 \; \theta_{23}  \cos^2 \theta_{13} \; 
 \sin^2 \frac{\Delta m^2_{23} L}{4 E_\nu} 
\label{todasprobs}
\end{eqnarray}

Notice that all the probabilities depend in the same way of $\Delta m^2_{atm}$. The dependence
with the angle $\theta_{13}$ is such that in the limit $\theta_{13} \rightarrow 0$ one
recovers the two-family oscillations formulae.

Precisely the fact that $\theta_{13}$ is small (the CHOOZ experiment\cite{chooz} has set a
limit $\sin^2 \theta_{13} < 0.05$), together with the strong mass hierarchy 
( $\Delta m^2_{atm} >> \Delta m^2_{sun}$) results in 
the solar and atmospheric oscillations approximately decoupling in 2-by-2 
mixing phenomena. A consequence of
this is that future solar experiments\cite{newsolar} will improve the knowledge in the solar parameters
$\Delta m^2_{12}, \theta_{12}$
while future atmospheric and long base line accelerator experiments\cite{futurelbl}
will improve the knowledge of the atmospheric
parameters $\Delta m^2_{23}, \theta_{23}$, but they can learn very little about
{\em (i)} $\theta_{13}$ which links the solar and atmospheric oscillations, 
{\em (ii)} the sign of $\Delta m^2_{23}$
(which specifies the neutrino mass spectrum) and {\em (iii)} the CP-odd phase
$\delta$. These topics are the almost exclusively realm of a neutrino factory.

If the solar solution lies in the LMA-MSW region then 
$\Delta m^2_{sun} \sim \Delta m^2_{atm}/10 - \Delta m^2_{atm}/100$, and the
approximation which leads to formulae \ref{todasprobs} is no longer valid for
sufficiently small values of $\theta_{13}$. Instead,
a good and simple approximation for the $\nu_e \raw \nu_\mu$ transition probability 
is obtained by expanding to second order in the small parameters, 
$\tetaot, \Delta_{12} / \Delta_{13}$ and $\Delta_{12} L$\cite{golden}:
\begin{eqnarray}
P_{\nu_ e\nu_\mu  ( \bar \nu_e \bar \nu_\mu ) } & =&  
s_{23}^2 \, \sin^2 2 \tetaot \, \sin^2 \left ( \frac{\delot \, L}{2} \right )  \nn \\
&+ & c_{23}^2 \, \sin^2 2 \theta_{12} \, \sin^2 \left( \frac{ \Delta_{12} \, L}{2} \right ) 
\nn \\
& + & \tilde J \, \cos \left ( \pm \delta - \frac{ \delot \, L}{2} \right ) \nn \\
& & \frac{ \Delta_{12} \, L}{2} \sin \left ( \frac{ \delot \, L}{2} \right ) \, , 
\label{vacexpand} 
\end{eqnarray}
where
\bea
\Delta_{ij} \equiv \frac{\Delta m^2_{ij} }{2 E_\nu} \, .
\eea
and
\bea
\tilde J \equiv c_{13} \, \sin 2 \theta_{12} \sin 2 \tetatt \sin 2 \tetaot 
\eea
is the combination of mixing angles appearing in the Jarlskog determinant. 

Notice that, according to equation \ref{vacexpand}, the CP-odd term is
proportional to $J$ (and therefore to the product of all the mixing angles), and
also to $\Delta m^2_{sum}$. Therefore any CP asymmetry will be suppressed by
the solar $\Delta m^2$ and mixing angle and will become too small to
be measurable if those parameters are too small, as would be the case
if the solar solution does not lie in the LMA-MSW region. Fortunately, recent
data from SuperKamiokande\cite{newSK,concha} favors precisely this region.
In the rest of this paper I will assume that nature is kind and the LMA-MSW solution
is indeed the true one. This expectations will be confirmed in a few years from
now, by forthcoming solar experiments\cite{newsolar}.
 
The above formulae are obtained assuming propagation in vacuum. However,
when $\nu$'s cross the earth, forward scattering amplitudes are different for
the different flavors:
\begin{small}
\begin{eqnarray}
M^2_\nu &  = &V_{NMS} \left(\matrix{m^2_1 & & \cr
 & m^2_2 & \cr
 &  & m^2_3 \cr}\right) V_{NMS}^\dagger \; \nn \\
 & + & \left(\matrix{\pm 2 E_\nu A & & \cr
 &  & \cr
 &  & \cr}\right) 
\end{eqnarray}
\end{small}
where  $A \equiv \sqrt{2} G_F n_e$ and
$n_e$ the ambient electron number density\cite{MSW}. 
The presence of matter modifies the transition probabilities which can be written
(for example for the $\nu_\mu \rightarrow \nu_e$ transitions) as:
\begin{eqnarray}
P_{\nu_e\nu_\mu({\bar \nu}_e {\bar \nu}_\mu)} & = 
& \sin^2 \theta_{23} \sin^2 2\; \theta_{13} \nn \\ 
& & \left( \frac{\Delta m^2_{23}}{B_\pm}\right)^2 \; \; 
\sin^2  B_{\pm} L 
\end{eqnarray}
which has the same form of the corresponding probability in vacuum 
(equations \ref{todasprobs})
substituting the mixing angle $\theta_{23}$ by an ``effective mixing angle''
$\sin^2 \theta_{23} \sin^2 2\; \theta_{13} 
\left( \frac{\Delta m^2_{23}}{B_\pm} \right)$ and the mass difference
$\Delta m^2_{atm}/4E_\nu$ by an ``effective mass difference'':
\begin{eqnarray} 
B_{\pm} &\equiv& 
((\Delta m^2_{23} \cos 2~\theta_{13} \pm 2 E_\nu A)^2 \nn \\ &+&
 (\Delta m^2_{23} \sin 2~\theta_{13})^2)^{1/2} 
\end{eqnarray}

Matter effects on Earth are important if $A$ is comparable to, or bigger than, 
$\Delta_{atm}$ for some neutrino energy, 
and if the distances traveled through the Earth 
are large enough for the probabilities to be in the non-linear 
region of the oscillation. 

For the Earth's crust, with density $\rho \sim 2.8 g/cm^3$ and roughly equal numbers of protons, 
neutrons and electrons, $A \sim 10^{-13} eV$. 
The typical neutrino energies we are considering are tens of GeVs.
For instance, for $E_\nu=30 GeV$ (the average $\bar\nu_e$ energy in the 
decay of $E_\mu=50 GeV$ muons) $A = 1.1 \times 10^{-4} ~eV^2/GeV$ 
$\sim \deltt$. This means that matter effects will be important at long distances. 
 Since the 
``effective mass'' $B_{\pm}$ is CP-odd, the net effect of matter is to
induce, at sufficiently large baselines a ``fake'' CP-asymmetry which hides
genuine CP-violation. 

\section{THE NEUTRINO FACTORY}
\label{ref:nufact}
\subsection{The Machine}
\begin{figure}[tbp]
\begin{center} \epsfig{file=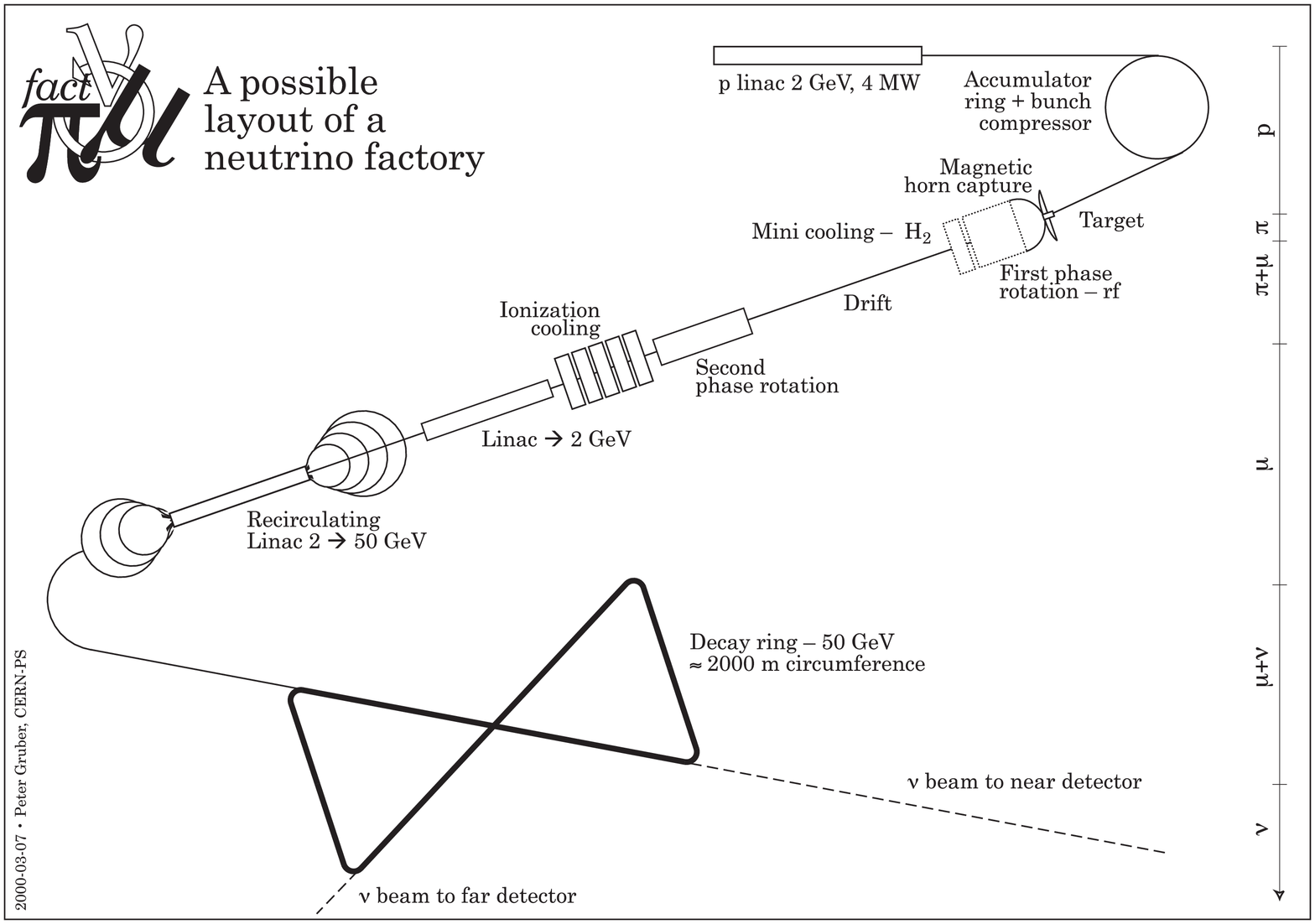,height=8cm, width=8cm, angle=0}
\end{center}
\caption{Generic Layout of a Neutrino Factory.}
\label{nufact}
\end{figure}

The generic layout of a neutrino factory  is shown in Fig.~\ref{nufact}.
A high power (4~$MW$) proton beam impinges on a target producing
pions which are collected and focused with a magnetic device (such as the 
magnetic horn depicted in the figure) and let decay in a drift space.
Next, the muon phase space is reduced (phase rotation, ionization cooling) and 
the muons are injected into a set of linacs which accelerate them up
to an energy of 50~$GeV$. Finally they are fed into 
a storage ring.
The muons decaying in the
straight sections of this ring produce a high
intensity, pure neutrino beam that points towards a neutrino detector
(a bow-tie design, such as the one shown in the figure allows two different
locations). By changing
the sign of the charge of the collected pions it is possible to
get the two conjugated neutrinos.

The design parameters of the neutrino factory have been extensively
discussed in the Lyon and Monterey workshops\cite{lyon,monterey}. 
The results discussed here were obtained assuming an integrated data
set of $10^{21}$ 
useful $\mu^+$ decays and $10^{21}$ useful $\mu^-$ 
decays, a  muon beam energy of 50~$GeV$, no polarization and a detector mass of 40~$kt$.

\subsection{Wrong sign muons}

As one can see in formula \ref{vacexpand}, in order to be sensitive to the parameters
$\tetaot$ and $\delta$ one must measure the transition 
probabilities involving $\nu_e$ and $\bar \nu_e$, in particular 
$\nu_e(\bar \nu_e) \raw \nu_\mu ( \bar \nu_\mu)$. The neutrino factory 
is unique in providing high energy and intense 
$\nu_e (\bar \nu_e)$ beams coming from positive (negative) muons. 
Since these beams contain also $\bar \nu_\mu (\nu_\mu)$ (but no 
$\nu_\mu (\bar \nu_\mu)$ as is the case for conventional
neutrino beams), the transitions of interest can be measured by 
searching for ``wrong-sign'' muons\cite{geer,dgh}, e.g., 
negative (positive) muons  
appearing in a (massive) detector with good muon charge identification 
capabilities, provided that the non-beam backgrounds (i.e, backgrounds
arising from the bulk of \numu\ and \nue\ charged and neutral current
events) to this signal can be kept sufficiently small. 
Notice that there are no other neutrino flavors in the beam,
unlike the case of conventional hadron beams which contain an irreducible 
contamination of other flavors due to the decay of kaons and opposite-sign
pions. 

\subsection{A Large Magnetic Detector for the Neutrino Factory}

\begin{figure}[htbp]
\begin{center}
\mbox{\epsfig{file=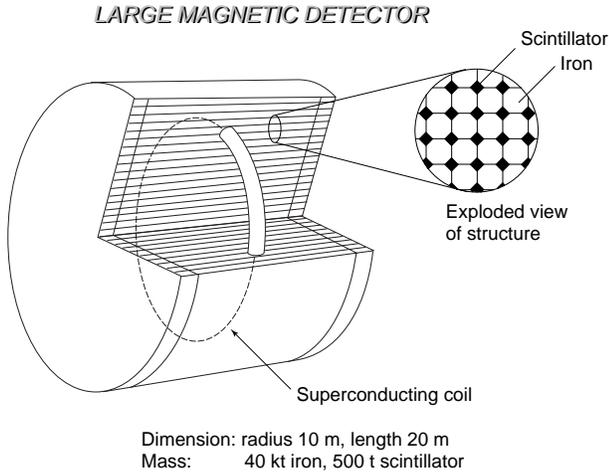,width=8cm}}
\end{center}
\caption{Sketch of the large magnetic detector for the 
        neutrino factory.}
\label{fig:detector}
\end{figure}

The detector proposed in\cite{cdg} is
shown in Fig.~\ref{fig:detector}. It is a large cylinder 
of 10~$m~$ radius and 20~$m~$ length, made of 6~$cm~$ thick iron rods 
interspersed with 2~$cm~$ thick scintillator rods built of 2~$cm~$ long
segments (light readout on both ends allows the determination of the 
spatial coordinate along the scintillator rod).  
Its mass is 40~$kt$. A super conducting coil generates a
solenoidal magnetic field of 1~$T~$ inside the iron. 
  
A neutrino traveling through the detector
sees a sandwich of iron and scintillator, with the $X,Y$ coordinates
being measured from the location of the scintillator rods and the $Z~$
coordinate being measured from their longitudinal segmentation. 

Neutrino interactions in such a detector have a clear signature. 
A CC \numu\ event is characterized by a muon, 
easily seen as a penetrating
track of typically several meters length, 
and a shower resulting from the interactions of the final-state hadrons.
A NC event, though, contains no penetrating
track and the length of the event is the length of the hadron shower in
iron, typically 
less than one meter. CC \nue\ events, on the other hand,
cannot be easily recognized since, with a detector of this coarse 
granularity, it is difficult to disentangle the prompt
electron from the hadronic shower on an event-by-event
basis. The performance of the  detector will be
similar to that of MINOS\cite{minos}. The main difference
lies in the mass which is one order of magnitude larger, and in
the smaller surface-to-volume ratio.

The potential backgrounds to the wrong-sign muon signal events are
NC events (as well as CC events in which the right-sign lepton is not
detected) in which a secondary negative muon arising from the decay of
$\pi^-, {\rm K}^-$ and ${\rm D}^-$ hadrons fakes the signal.
The discrimination of these backgrounds 
is based on the fact that
the muon produced in a CC {\em signal} event  
is harder and more isolated from the hadron shower axis 
than the one produced from hadron decay in background events. 
Accordingly, in\cite{cdg} an analysis is performed
based on the momentum of the muon, $p_\mu$, and 
a variable
measuring the isolation of the muon from the hadron shower axis, 
$q_{\rm t} = p_\mu \sin \theta$.

\begin{figure}[htbp]
\begin{center}
\vspace*{-2cm}
\mbox{
\epsfig{file=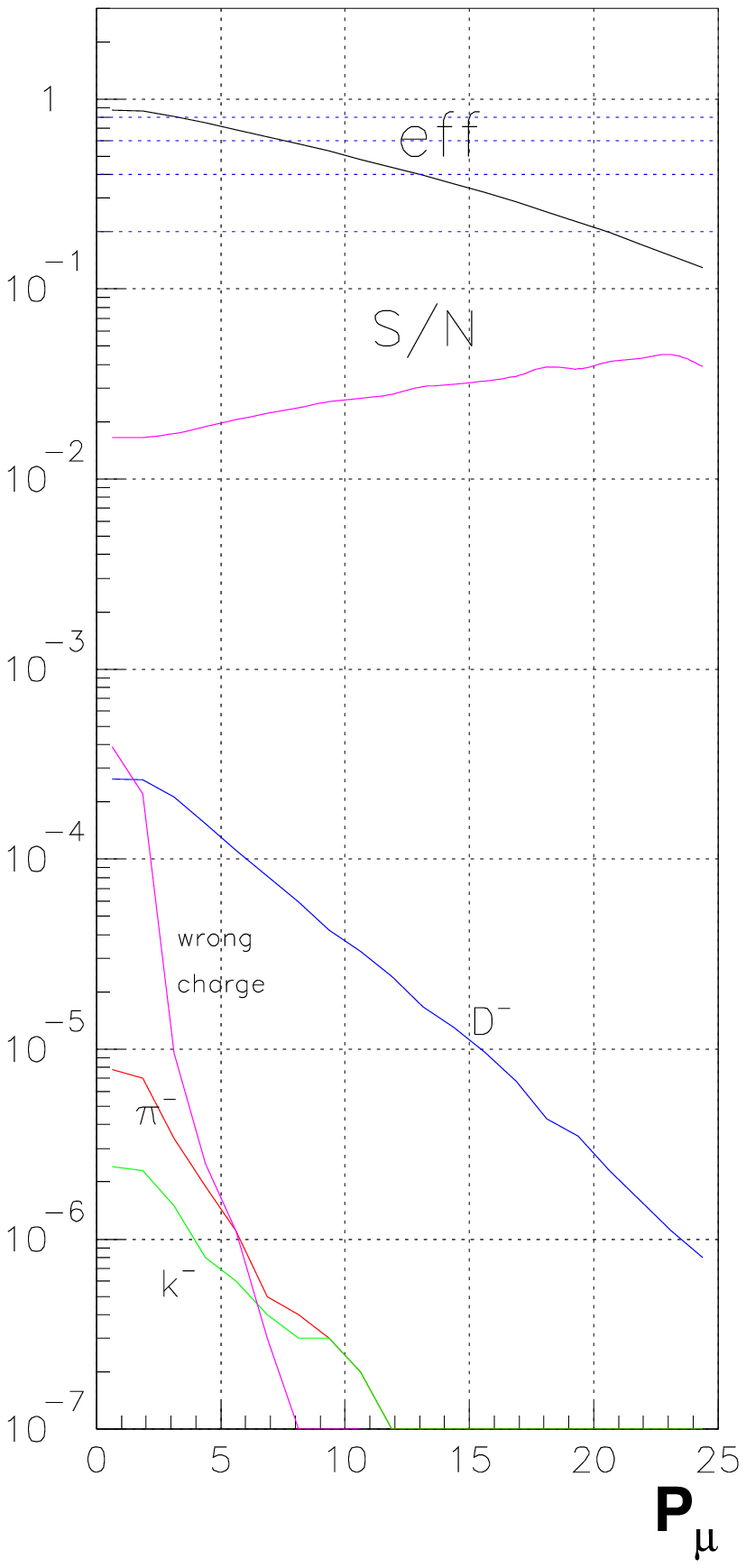,height=7cm,width=7cm}
}
\mbox{
\epsfig{file=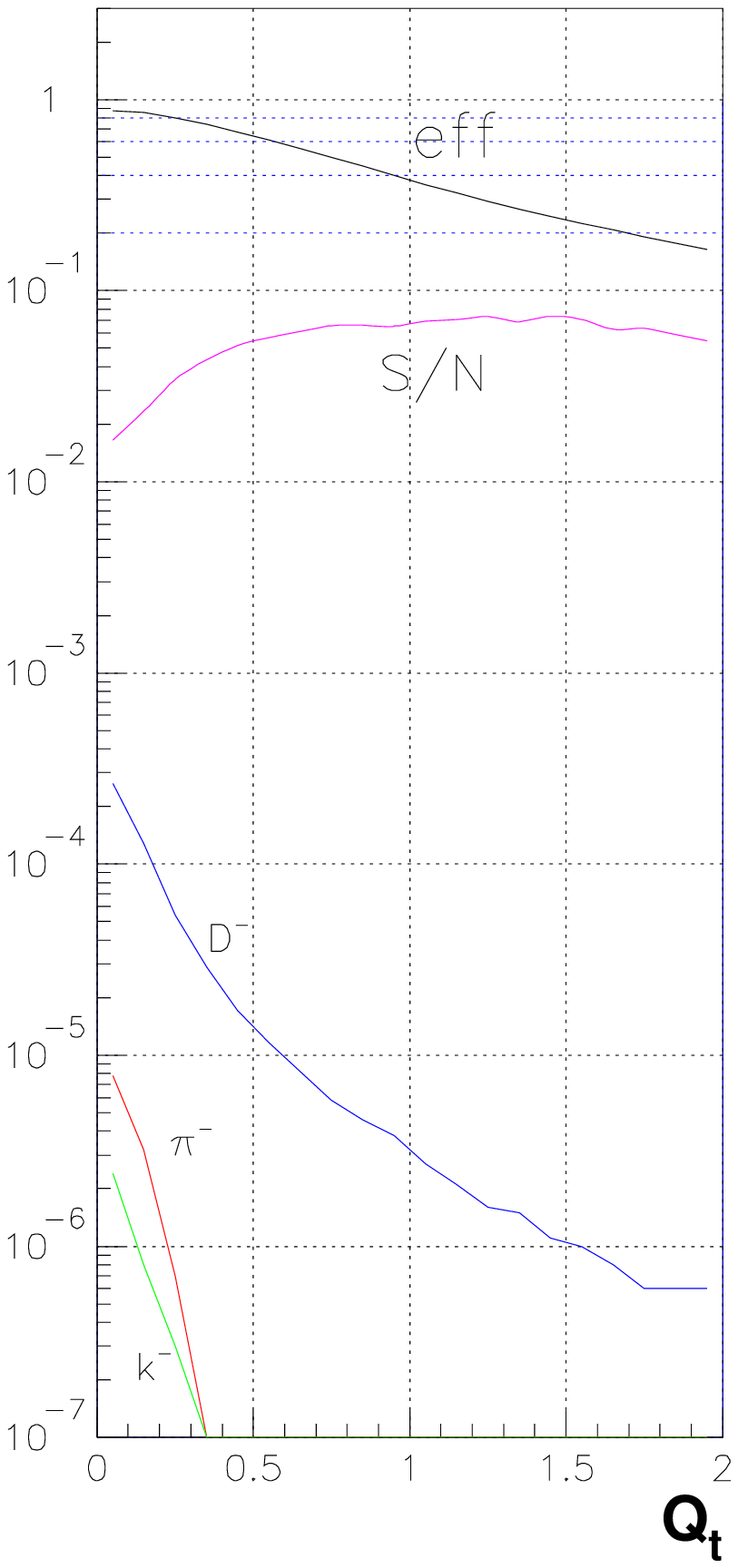,height=7cm,width=7cm}
}
\end{center}
\caption{Wrong-sign muon efficiency and fractional 
backgrounds for \numubar\ CC  events, as a
function of  
$p_\mu$ or $q_{\rm t}$, for 
a neutrino beam
originating from 50 $~GeV/c~$ $\muplus$ decays.}
\label{fig:ana1}
\end{figure}

An example of the rejection power of this analysis can be seen in 
Fig.~\ref{fig:ana1} which
shows the efficiency for signal detection as well as
the fractional backgrounds due to ``right'' sign charged currents,
in which the ``right sign muon'' is lost. Two independent plots
are shown, one  
as a
function of the cut on $p_\mu$ and the oter
as a function  of the cut on $q_{\rm t}$.
Also shown is the ratio $S/N = \varepsilon_{\rm s}/\sigma_{\rm b}$, where
$\varepsilon_{\rm s}$ is the signal selection efficiency and
$\sigma_{\rm b}$ is the error in the number of  
background events which survive the cuts. 
Muons from
charmed-hadron decays constitute the main background from 
\numubar\ CC events. Overall, a reduction of the background
at the level of $10^{-6}$ seems achievable. For an extensive
discussion I refer to \cite{cdg}.

\section{MEASUREMENT OF THE CP VIOLATION PHASE}
\label{ref:cp}
\begin{figure}[htbp]
\begin{center}
\epsfig{file=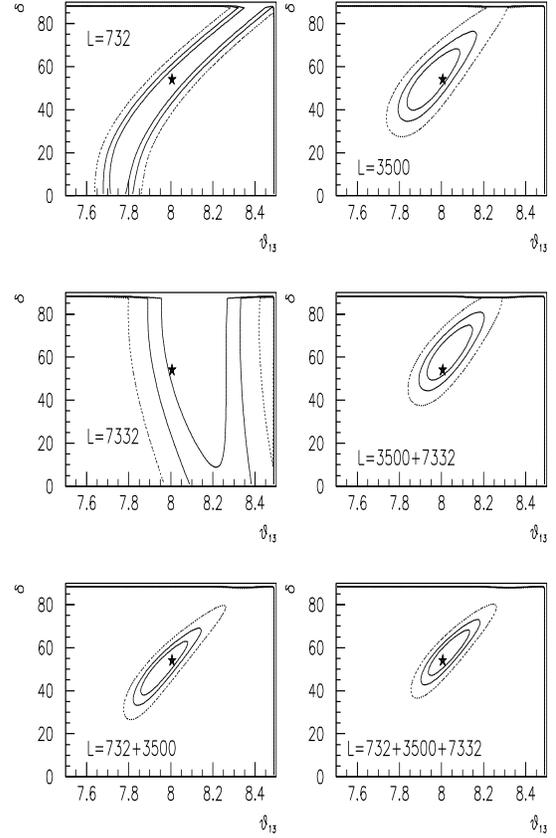,width=8cm,height=12cm} 
\end{center}
\caption{ 68.5, 90, 99 \% CL contours resulting from a $\chi^2$ fit of  
$\tetaot$ and $\delta$. The parameters used to generate the ``data'' are 
depicted by a star and the baseline(s) which is used for the fit indicated 
in each plot. Statistical errors, backgrounds and efficiencies 
are included.}
\label{fig:54}
\end{figure}

We are now in position to show how the CP violation phase could be measured
in the neutrino factory. As pointed out before,
I will assume that the solar solution is in the currently
favored range LMA-MSW. Also, to simplify the discussion
fixed values of the atmospheric parameters are used in this section, 
$\Delta m_{23}^2 = 2.8\times 10^{-3} eV^2$ and maximal mixing,
$\tetatt=45^\circ$. 

Let us start discussing the measurement of the 
CP phase $\delta$ versus $\tetaot$.
Consider first the upper solar mass range allowed by the LMA-MSW solution: 
$\Delta m_{12}^2=10^{-4}$ eV$^2$. Fig.~\ref{fig:54} shows the confidence 
level contours for a simultaneous fit of $\tetaot$ and $\delta$, for 
Montecarlo generated data (including detector response) 
corresponding to $\tetaot=8^\circ$, $\delta=54^\circ$. The results include
statistical errors as well as those due to background subtraction. Detection efficiencies
are also taken into account. Genuine CP violation
is separated from the fake CP violation induced by matter effects taking advantage
of the different dependency on energy and base line (see\cite{golden} for a detailed discussion).

Notice that at ``short'' distances (i.e., 700 Km)
the correlation between $\delta$ and $\tetaot$ is very large. 
The phase $\delta$ is not measurable and this indetermination 
induces a rather large error on the angle $\tetaot$. However, at the 
intermediate baseline of 3500 km the two parameters can be disentangled 
and measured. At the largest baseline, the sensitivity to $\delta$ is lost 
and the precision in $\tetaot$ becomes worse due to the smaller statistics. 
The combination of the results for 3500 km with that for any one of the 
other distances improves the fit, although not in a dramatic way. However,
also from the point of view of understanding systematics I believe that two base lines
are preferred. Notice that
a CERN-based neutrino factory could choose a ``short base line'' experiment
located in the  Gran Sasso laboratory in Italy, at a distance of 
about 700 Km\footnote{This is an ideal location for one experiment since the current
generation of neutrino experiments will start to take data there
in a few years from now.}. The ``long base line'' experiment must be located,
as discussed at about 3000 Km. Possible locations, with good potential underground
sites\footnote{The detector(s) discussed must be deep underground to reduce the
huge flux of cosmic rays.} exist in Spain (in La Palma, one of the Canary islands)
if the second beam shoots south or in Norway and/or Finland if shooting north.

\begin{figure}[htbp]
\begin{center}
\epsfig{file=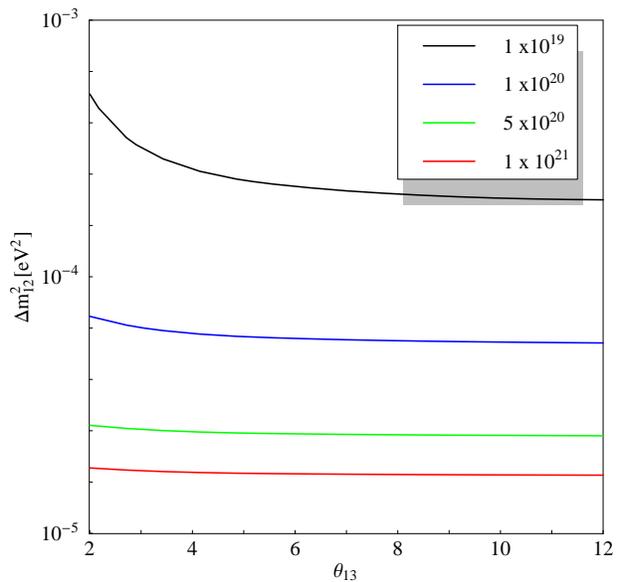,width=9cm}
\end{center} 
\caption{ Lower limit in $\Delta m_{12}^2$ at which a maximal CP phase 
($90^\circ$) can be distinguished from a vanishing phase at 99\% CL, as a 
function of $\tetaot$ at $L = $ 3500km and for different numbers of 
useful muons. Background errors and efficiencies are included.}
\label{fig:limdm12}
\end{figure}

The sensitivity to CP-violation decreases linearly with $\Delta m_{12}^2$.
At the central value allowed by the LMA-MSW solution, $\Delta m_{12}^2 = 
5\times 10^{-5} eV^2$, CP-violation can still be discovered, 
while for $\Delta m_{12}^2 = 1 \times 10^{-5} eV^2$, the sensitivity to 
CP-violation is lost with the experimental set-up used. 
We have quantified what is the minimum value of $\Delta m^2_{12}$ for which 
a maximal CP-odd phase, $\delta = 90^\circ$, can be distinguished at 99\% CL 
from $\delta = 0^\circ$. The result is shown in Fig.~\ref{fig:limdm12}: 
$\Delta m^2_{12} > 2 \times 10^{-5} eV^2$, with very small dependence 
on $\tetaot$, in the range considered. 

For an extensive discussion I refer to \cite{golden,romanino,otherCP}.

\section*{Acknowledgments}
Much of the quantitative results presented in this paper are part of studies made in
collaboration with A. Cervera, F.Dydak, A. Donini, B. Gavela, P.Hernandez, O. Mena and  
S. Rigolin. 

In spite of the fact that the birth of my daugther Irene in the same week of CP2000 
restricted my participation in this wonderful conference I was nonetheless able to contribute
thanks to the kindness, patience and understanding of the organizers to which I would like to
express my most sincere acknowledgments.

\end{document}